\documentclass[12pt]{iopart}
\usepackage{iopams}
\usepackage{dsfont}
\usepackage[dvips]{graphicx}

\newcommand{\ri}{\mbox{$\rm i$}}
\newcommand{\rd}{\mbox{$\rm d$}}

\begin{document}

\title[Nonlinearity and constrained quantum motion]
{Nonlinearity and constrained quantum motion}

\author[D~C~Brody, A~C~T~Gustavsson, and L~P~Hughston]
{Dorje~C.~Brody${}^1$, Anna~C.~T.~Gustavsson${}^2$, and
Lane~P.~Hughston${}^1$}

\address{${}^1$ Department of Mathematics, Imperial College
London, London SW7 2AZ, UK}

\address{${}^2$ Blackett Laboratory, Imperial
College London, London SW7 2AZ, UK}

\begin{abstract}
The dynamical equation satisfied by the density matrix, when a quantum
system is subjected to one or more constraints arising from
conserved quantities, is derived. The resulting nonlinear motion of the density matrix has the property that the
evolution is independent of the specific composition of the pure-state mixture generating the initial state of the system.
\end{abstract}

\submitto{\JPA}
%
%\pacs{05.30.-d, 02.40.Yy, 03.65.Yz}
%\maketitle
\vspace{0.4cm}

%\section{Introduction}
%\label{sec:1}

A nonlinear generalisation of quantum mechanics was proposed
by Mielnik~\cite{mielnik,mielnik1}, Kibble~\cite{kibble}, and
Weinberg \cite{wein} as an alternative to the linear evolution
governed by the Schr\"odinger equation. The idea of this theory
is to replace the linear Hamiltonian operator
generating the evolution of the state vector with a state-dependent operator
in such a way that the norm of the state vector remains constant. The properties of such ``nonlinear observables''
were subsequently studied in detail by Weinberg \cite{wein2}.

Following Weinberg's analysis, it was argued by Gisin
\cite{gisin,gisin2} and Polchinski \cite{polch} that the existence
of dynamical nonlinearities in quantum mechanics might lead to
undesirable physical features, such as the possibility of superluminal
EPR communication. Indeed, it is
often held now that the main issue associated
with dynamics of the Mielnik-Kibble-Weinberg (MKW) type
is that the evolution of a density matrix in general depends on the
specific choice of pure-state mixture underlying the initial density
matrix. Thus by observing the dynamics
one might be able to determine the particular choice of mixture
associated with the initial density matrix (Haag \&
Bannier \cite{haag}).

Another potentially problematic aspect of nonlinear quantum
mechanics was suggested by Peres \cite{peres},
who provided an example where the von Neumann entropy
decreases in time. Tight experimental bounds on the deviation
away from linear (unitary) evolution law have been found~\cite{boll},
while criticism of the MKW theory has been strengthened further by
Mielnik~\cite{mielnik2}. As a consequence of these studies, one might conclude that a consensus has emerged to the effect that
nonlinear quantum mechanics of the MKW type must be ruled out
on physical grounds. 

The purpose of the present paper is to report a generalisation of the
MKW theory that circumvents some of these issues. The nonlinear dynamics that we propose emerges
from the consideration of constrained quantum motions. We shall
derive the dynamical equation satisfied by a density matrix when
unitarity is compromised by the existence of one or more constraints.
We demonstrate that although the resulting equations of motion
are nonlinear, the associated evolution is autonomous and hence independent of the choice
of initial mixture. We
conclude that the nonlinearities arising from the type of constraints
considered here can be regarded as representing a viable step towards an
acceptable generalisation of the unitary evolution of quantum mechanics.

Constrained motions appear not infrequently in the general study of
dynamical systems~\cite{rund}. A systematic investigation of
constrained motions in classical mechanics from a Hamiltonian point
of view was carried out
by Dirac~\cite{Dirac-50,Dirac-58}. In classical mechanics, the
evolution is governed by a symplectic flow on phase
space. The idea of Dirac, in essence, is to find the induced
symplectic structure on the constraint surface in phase space, and
to use this to characterise the dynamics. Dirac's approach has
recently been applied in the quantum context to obtain the
constrained dynamical equations satisfied by pure
states~\cite{Buric,BGH1,Corichi} in various examples. In this paper we apply similar techniques to derive constrained equations of motion for mixed states, and to show that in the case of a pure-state density matrix the equations
reduce to the results obtained previously. We consider in particular the special case where the constraints are
given by the conservation of the expectation values
of a family of mutually incompatible
observables $\{{\hat\Phi}^k\}$ ($k=1,\ldots,N$) such that $[{\hat
H},{\hat\Phi}^k]\neq0$ for all $k$, where ${\hat H}$ is the Hamiltonian. Given
an initial mixed-state density matrix ${\hat\rho}_0$ there are
infinitely many different mixtures of pure states that give rise to ${\hat\rho}_0$. In our scheme, nevertheless, the resulting
nonlinear evolution for the density matrix is independent of the
specific choice of mixture. We show also, for one version of the dynamics we consider, that the von Neumann
entropy is a constant of the motion, and hence that the criticism of Peres
does not apply to the constrained dynamics arising in that case.

Let us now present our analysis. As remarked above, the constraints that arise most naturally in the case of constrained
motions of density matrices involve the conservation of a family of
observables $\{{\hat\Phi}^k\}_{k=1,\ldots,N}$ so that
\begin{eqnarray}
{\rm tr}({\hat\rho}{\hat\Phi}^k) = c^k \label{eq:1}
\end{eqnarray}
for all $k$, where $\{c^k\}$ are constants. For the moment we shall
assume that $N$ is even. The constraints are assumed to be non-redundant
in the sense that $[{\hat\Phi}^k,{\hat\Phi}^l]\neq0$ for $k\neq l$
and $[{\hat H}, {\hat\Phi}^k]\neq0$ for all $k$. We shall impose
(\ref{eq:1}) by use of Lagrange multipliers, and write
\begin{eqnarray}
\frac{\rd \hat{\rho}}{\rd t} = \ri [\hat{\rho},\hat{H}] - \ri
\sum_{k=1}^N \lambda_k [\hat{\rho},\hat{\Phi}^k]  \label{eq:2}
\end{eqnarray}
for the proposed dynamics of the density matrix. Here $\{\lambda_k\}$, $k=1,\ldots,N$, are the
Lagrange multipliers conjugate to the constraints $\{c^k\}$. In what follows
we employ the usual convention that repeated indices are
summed.

The object is to impose the constraints and derive explicit formulae (given by nonlinear functionals of $\hat\rho$) for the Lagrange multipliers. To derive the $\{\lambda_k\}$ we consider the equation
\begin{eqnarray}
\frac{\rd}{\rd t}\, {\rm tr}({\hat\rho}{\hat\Phi}^k) = {\rm tr}
\left( \frac{\rd \hat{\rho}}{\rd t}\,\hat{\Phi}^k \right)=0.
\label{eq:3}
\end{eqnarray}
Substituting (\ref{eq:2}) in (\ref{eq:3}) gives
\begin{eqnarray}
{\rm tr}([\hat{\rho},\hat{H}]\hat{\Phi}^k]) - \lambda_j {\rm tr}
([\hat{\rho},\hat{\Phi}^j]\hat{\Phi}^k) = 0. \label{eq:4}
\end{eqnarray}
By the cyclic property of the trace operation, which says $\tr(\hat{A}\hat{B}\hat{C})=\tr(\hat{C}\hat{A}\hat{B})$, this equation can be
rewritten in the form
\begin{eqnarray}
{\rm tr}(\hat{\rho} [\hat{H},\hat{\Phi}^k]) - \lambda_j {\rm tr}
(\hat{\rho} [\hat{\Phi}^j,\hat{\Phi}^k]) = 0. \label{eq:5}
\end{eqnarray}
To proceed we define the antisymmetric matrix $w^{jk}$ by
\begin{eqnarray}
w^{jk}={\rm tr}(\hat{\rho} [\hat{\Phi}^j,\hat{\Phi}^k]).
\label{eq:6}
\end{eqnarray}
Provided that $w^{jk}$ is nonsingular, which we assume holds at least initially, let us write $w_{ij}$ for its inverse so that
$w_{ij}w^{jk}=\delta^k_i$. Then the equation reads
\begin{eqnarray}
\tr(\hat\rho[\hat{H},\hat\Phi^k]) = \lambda_j w^{jk}
\end{eqnarray}
and we can solve (\ref{eq:5}) for
the Lagrange multipliers to obtain
\begin{eqnarray}
\lambda_k = w_{jk} {\rm tr}(\hat{\rho} [\hat{H},\hat{\Phi}^j]).
\label{eq:7}
\end{eqnarray}
Substitution of this expression into (\ref{eq:2}) then gives
\begin{eqnarray}
\frac{\rd \hat{\rho}}{\rd t} = \ri [\hat{\rho},\hat{H}] - \ri w_{jk}
{\rm tr}(\hat{\rho} [\hat{H},\hat{\Phi}^j])
[\hat{\rho},\hat{\Phi}^k]. \label{eq:8}
\end{eqnarray}
This is the nonlinear equation of motion satisfied by the density matrix when it is
subject to an even number of constraints of the form (\ref{eq:1}).

Our objective next is to show that in the case of a pure-state
density matrix the evolution equation (\ref{eq:8}) reduces in effect to the
nonlinear Schr\"odinger equation obtained in reference \cite{BGH1}. In particular, suppose that $\hat\rho$ is a time-dependent pure-state density matrix of the form
\begin{eqnarray}\label{eq-rho-pure}
\hat\rho = \frac{|\psi\rangle\langle\psi|}{\langle\psi|\psi\rangle}
\end{eqnarray}
for some state vector $|\psi\rangle$, not necessarily normalised. Then we see that
\begin{eqnarray}
\frac{\rd \hat\rho}{\rd t}  =  \frac{|\dot\psi\rangle\langle\psi|}{\langle\psi|\psi\rangle} + \frac{|\psi\rangle\langle\dot\psi|}{\langle\psi|\psi\rangle}  - \frac{|\psi\rangle\langle\psi|}{\langle\psi|\psi\rangle^2}\left( \langle\dot\psi|\psi\rangle+\langle\psi|\dot\psi\rangle\right),
\end{eqnarray}
and hence that
\begin{eqnarray}
\frac{\rd \hat\rho}{\rd t} |\psi\rangle = |\dot\psi\rangle - \frac{\langle\psi|\dot\psi\rangle}{\langle\psi|\psi\rangle}|\psi\rangle.
\end{eqnarray}
Therefore, for example, if $\hat\rho$ is assumed to satisfy the von Neumann equation
\begin{eqnarray}
\frac{\rd \hat\rho}{\rd t} = \ri [\hat\rho,\hat{H}],
\end{eqnarray}
we can deduce that $|\psi\rangle$ satisfies the so-called projective Schr\"odinger equation
\begin{eqnarray}\label{eq-star}
|\dot\psi\rangle - \frac{\langle\psi|\dot\psi\rangle}{\langle\psi|\psi\rangle}|\psi\rangle = -\ri(\hat{H} - \langle\hat{H}\rangle)|\psi\rangle,
\end{eqnarray}
where $\langle\hat{H}\rangle$ denotes the expectation of the Hamiltonian:
\begin{eqnarray}
\langle\hat{H}\rangle = \frac{\langle\psi|\hat{H}|\psi\rangle}{\langle\psi|\psi\rangle}.
\end{eqnarray}
The projective Schr\"odinger equation is essentially a slightly ``weakened'' form of the ``full'' Schr\"odinger equation
\begin{eqnarray}\label{eq-star2}
|\dot\psi\rangle = -\ri\hat{H}|\psi\rangle,
\end{eqnarray}
with the ``unphysical'' component of $|\dot\psi\rangle$ in the direction of $|\psi\rangle$ removed. Clearly (\ref{eq-star2}) implies (\ref{eq-star}). It is worth noting, however, that although (\ref{eq-star2}) is linear, the associated projective equation (\ref{eq-star}), which embodies the physical content of the Schr\"odinger equation, is nonlinear.

In the case of a constrained quantum system satisfying (\ref{eq:2}), essentially the same line of argument applies, and a short calculation shows that if $\hat\rho$ is a pure-state density matrix then
\begin{eqnarray}
|\dot\psi\rangle - \frac{\langle\psi|\dot\psi\rangle}{\langle\psi|\psi\rangle}|\psi\rangle = -\ri(\hat{H} - \langle\hat{H}\rangle)|\psi\rangle + \ri\lambda_k(\hat\Phi^k - \langle\hat\Phi^k\rangle)|\psi\rangle,
\end{eqnarray}
with
\begin{eqnarray}
\langle\hat\Phi^k\rangle = \frac{\langle\psi|\hat\Phi^k|\psi\rangle}{\langle\psi|\psi\rangle},
\end{eqnarray}
and where by analogy with (\ref{eq:6}) and (\ref{eq:7}) we have defined
\begin{eqnarray}
\lambda_k = w_{jk} \frac{\langle\psi|[\hat{H},\hat\Phi^j]|\psi\rangle}{\langle\psi|\psi\rangle},
\end{eqnarray}
where $w_{jk}$ is the inverse of the matrix
\begin{eqnarray}
w^{ij}=\frac{\langle\psi|[\hat\Phi^i,\hat\Phi^j]|\psi\rangle}{\langle\psi|\psi\rangle}.
\end{eqnarray}
This result is indeed consistent with the
equation derived in reference \cite{BGH1}, and thus we are able to conclude that the dynamical equation (\ref{eq:8})
is a natural mixed-state generalisation of the constrained equation of
motion for pure states.

We remark, as we discuss in more detail below, that the motion generated by the constrained
equation (\ref{eq:8}) depends only on the initial density matrix, and not
on the choice of mixture leading to that matrix. This follows simply from the fact that (\ref{eq:8}) is autonomous in $\hat\rho$. In addition, we can verify, by
use of the cyclic property of the trace operation, that the von Neumann
entropy, defined by
\begin{eqnarray}
S=-{\rm tr}({\hat\rho}\ln{\hat\rho}),
\end{eqnarray}
is a constant of motion. The proof is as follows. First we note that if we let $p_n(t)$ denote the components of $\hat\rho$ along the diagonal in a Hilbert space basis with respect to which $\hat\rho$ is diagonalised, we have
\begin{eqnarray}
\frac{\rd S}{\rd t} & = & -\frac{\rd}{\rd t} \tr(\hat\rho \ln \hat\rho) \nonumber\\
{} & = & -\frac{\rd}{\rd t} \sum_n p_n \ln p_n \nonumber\\
{} & = & - \sum_n \dot{p}_n \ln p_n \nonumber\\
{} & = & - \tr\left(\frac{\rd \hat\rho}{\rd t} \ln \hat\rho\right).
\end{eqnarray}
Then by use of the dynamical equation (\ref{eq:8}) for the density matrix we obtain
\begin{eqnarray}
\frac{\rd S}{\rd t} & = & -\ri \tr\left([\hat{H} - \sum_k\lambda_k \hat\Phi^k,\hat\rho]\, \ln\hat\rho\right)\nonumber\\
{} & = & -\ri \tr\left([\ln\hat\rho,\hat\rho]\left(\hat{H} - \sum_k\lambda_k\hat\Phi^k\right)\right)\nonumber\\
{} & = & 0
\end{eqnarray}
by virtue of the cyclic property of the trace, and the fact that $\hat\rho$ and $\ln \hat\rho$ commute. We can therefore regard (\ref{eq:8}) as
representing a plausible candidate for an acceptable extension of the standard unitary dynamics defined by the von Neumann equation.

We observe that the objections
originally raised against nonlinear extensions of quantum
mechanics were based implicitly on the essentially mistaken premise that the dynamics of a
general mixed-state density matrix can and must be deduced
from the dynamics of pure states. More precisely, if
the initial density matrix ${\hat\rho}(0)$ happens to take a
decomposition of the form
\begin{eqnarray}
{\hat\rho}(0) = \sum_n p_n {\hat\Pi}_n(0),
\end{eqnarray}
where $\{{\hat\Pi}_n(0)\}_{n=1,\ldots}$ are normalised projection
operators onto a set of pure states $\{|\psi_n\rangle_0\}_{n=1,\ldots}$,
then it was \emph{assumed} that the subsequent dynamics of the density
matrix would have to be of the linear form
\begin{eqnarray}
{\hat\rho}(t) = \sum_n p_n {\hat\Pi}_n(t).  \label{eq:12}
\end{eqnarray}
The point is that in his original analysis Gisin \cite{gisin,gisin2} had no way to deduce the dynamics of the density matrix except to regard it as following from the dynamics of pure states. From a modern perspective, however, we can take essentially the opposite view, and regard the density matrix
as ``fundamentally'' representing the state of the system, from which
properties of pure states can be deduced as special cases. Hence, in particular, there is no reason to suppose that the dynamics of $\hat\rho$ can or should be deduced, linearly, from the dynamics of a set of hypothetical ensemble constituents.

So far we have considered the case for which the number $N$ of
conserved observables is even. If $N$ is odd, then the antisymmetric
matrix $w^{jk}$ defined in (\ref{eq:6}) is singular and cannot
be inverted. Thus, to obtain a system of constrained equations of motion that
are applicable to the case for which $N$ is odd, we need to modify the
foregoing analysis. The idea here is to replace the commutator in
(\ref{eq:2}) by a symmetric product of the form
\begin{eqnarray}
\frac{\rd \hat{\rho}}{\rd t} = \ri [\hat{\rho},\hat{H}] - \lambda_k
\left( \{\hat{\rho},\hat{\Phi}^k\} - 2 {\rm tr}
(\hat{\rho}\hat{\Phi}^k) \hat{\rho} \right), \label{eq:x1}
\end{eqnarray}
where $\{\hat{\rho},\hat{\Phi}^k\}=\hat{\rho}\hat{\Phi}^k +
\hat{\Phi}^k\hat{\rho}$ denotes the anticommutator and where the
$\{\lambda_k\}$ comprise a set of Lagrange multipliers chosen to ensure the
constraints (\ref{eq:1}), where $N$ need not be even. The plan is to
circumvent the problem of the lack of invertibility of $w^{jk}$
arising from the antisymmetric feature of the commutator by replacing
it with a symmetric anticommutator $\{\hat{\rho},\hat{\Phi}^k\}$. The
additional trace term on the right side of (\ref{eq:x1}) is then to
ensure conservation of the total probability, so that
\begin{eqnarray}
\frac{\rd ({\rm tr}{\hat\rho})}{\rd t} =0,
\end{eqnarray}
which follows at once from equation (\ref{eq:x1}).

As before, we determine the Lagrange multipliers by considering the
relation (\ref{eq:3}). Substituting (\ref{eq:x1}) in (\ref{eq:3}), and using
the cyclic property of the trace, we deduce that
\begin{eqnarray}
{\rm tr}(\hat{\rho}[\hat{H},\hat{\Phi}^k]) =-\ri \lambda_j \left(
{\rm tr}(\hat{\rho}\{\hat{\Phi}^j,\hat{\Phi}^k\} )\!-\! 2{\rm tr}
(\hat{\rho}\hat{\Phi}^j){\rm tr}(\hat{\rho}\hat{\Phi}^k) \right).
\label{eq:x3}
\end{eqnarray}
To solve (\ref{eq:x3}) for $\lambda_j$  we define the symmetric covariance
matrix
\begin{eqnarray}
m^{jk}={\rm tr}(\hat{\rho}\{\hat{\Phi}^j,\hat{\Phi}^k\} ) - 2{\rm
tr}(\hat{\rho}\hat{\Phi}^j){\rm tr}(\hat{\rho}\hat{\Phi}^k).
\label{eq:x4}
\end{eqnarray}
Again, if $m^{jk}$ is nonsingular, we can define its inverse
$m_{ij}$ that satisfies $m_{ij}m^{jk}=\delta^i_k$. In this case, we
obtain the following expression for the Lagrange multipliers:
\begin{eqnarray}
\lambda_j=\ri m_{jk} {\rm tr} \left( \hat{\rho} [\hat{H},
\hat{\Phi}^k]\right) \label{eq:x5}
\end{eqnarray}
The invertibility of the matrix $m^{jk}$ is ensured by the
positive-definiteness of the covariance matrix, which holds if
${\hat\rho}$ is itself nonsingular. Therefore, under this assumption the generalised
equation of motion becomes
\begin{eqnarray}
\frac{\rd \hat{\rho}}{\rd t} = \ri [\hat{\rho},\hat{H}] - \ri m_{jk}
{\rm tr}\left(\hat{\rho}[\hat{H},\hat{\Phi}^k]\right)\! \left(
\{\hat{\rho},\hat{\Phi}^j\}\! -\! 2 {\rm tr}(\hat{\rho}\hat{\Phi}^j)
\hat{\rho} \right). \label{eq:17}
\end{eqnarray}

\begin{figure}
\begin{center}
  \includegraphics[scale=0.65]{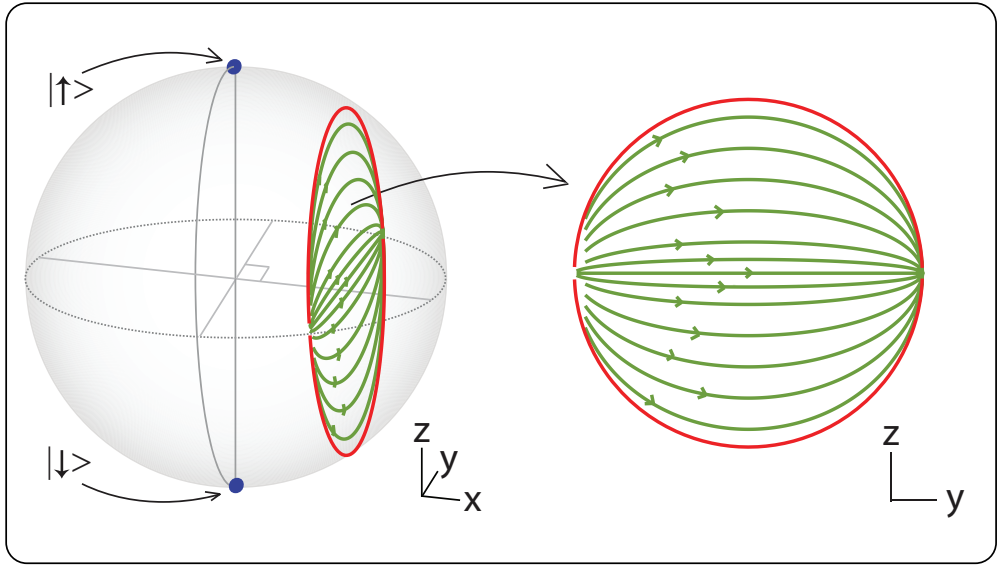}
  \caption{\textit{Constrained dynamics for mixed states in the
  case of a spin-$\frac{1}{2}$ system}. The state space is a ``Bloch
  ball'' ${\mathcal B}$ of unit radius; pure states form the
  boundary (surface) of ${\mathcal B}$ and interior points
  represent mixed states. The Hamiltonian of the system
  is given by ${\hat H}={\hat\sigma}_z$. Hence under unitarity motion
  the evolution generates a rigid rotation of ${\mathcal B}$
  around the $z$-axis. We constrain the motion by fixing
  the expectation of $\hat\sigma_x$; the resulting constraint surface
  is a slice of ${\mathcal B}$ perpendicular to the $x$-axis, as
  indicated above. Irrespective of the initial condition, the motion
  converges to the pure state for which $\langle{\hat\sigma}_z
  \rangle=0$ asymptotically. The integral curves associated
  with the dynamical equation are shown in the figure.
  \label{fig:1}
  }
\end{center}
\end{figure}

\emph{Example:} As an illustration let us consider the case of a
spin-$\frac{1}{2}$ system for which ${\hat H}={\hat\sigma}_z$ and the single
constraint observable is chosen to be
${\hat\Phi}={\hat\sigma}_x$. The resulting motion for the
density matrix is shown in Figure~\ref{fig:1}. The space of density matrices in this case is a ``Bloch ball'' $\mathcal B$ of unit radius. The mixed states correspond to the interior points of $\mathcal B$ and the pure states form the surface of $\mathcal B$. The chosen Hamiltonian would in the unconstrained case generate a rigid rotation around the $z$-axis. As we constrain the motion of the system by imposing the condition that the expectation of $\hat\sigma_x$ must be conserved, the resulting constraint surface corresponds to a cross-section of $\mathcal B$ at $x=x_0$, where $\tr(\hat\rho\hat\sigma_x)=x_0$. From (\ref{eq:17}) we deduce the equation of motion for the system, and find, as is shown in Figure~\ref{fig:1}, that when the initial state is given by a pure state density matrix, the state of the system remains pure as it evolves. In this case, the example reduces to the case considered in reference \cite{BGH2}. The mixed state evolution trajectories are also shown for a choice of initial states in the figure. The equator on the surface of $\mathcal{B}$ corresponds to a set of fixed states. Hence a state that initially lies at the point to the far left of the cross-section in Figure~\ref{fig:1} remains fixed, and all other states evolve asymptotically towards the fixed point to the right in the figure.

%%%%%%%%%%%%%%%%%%%%%%%%%%%
%%%%%%%%%%%%%%%%%%%%%%%%%%%%

Returning now to (\ref{eq:17}), we note that in the special case for which ${\hat\rho}$ is a pure-state density-matrix of the form
(\ref{eq-rho-pure}) we are able to deduce the nonlinear projective Schr\"odinger equation
satisfied by the state vector $|\psi\rangle$ that is applicable to
both even and odd number of conserved observables. This is given by
\begin{eqnarray}
|\dot\psi\rangle - \frac{\langle\psi|\dot\psi\rangle}{\langle\psi|\psi\rangle}|\psi\rangle =  -\ri (\hat{H}-\langle{\hat H}\rangle)
|\psi\rangle - \lambda_{k}(\hat{\Phi}^k-\langle{\hat\Phi}^k\rangle)
|\psi\rangle,  \label{eq:18}
\end{eqnarray}
which is consistent with the result obtained in reference \cite{BGH2}. Therefore, (\ref{eq:17}) constitutes a
natural generalisation of the result of \cite{BGH2} to the case of general density matrices.

It is interesting to observe that, unlike the motion
of (\ref{eq:8}), the motion of (\ref{eq:17}), which is applicable
to any number of constraints, does not necessarily preserve
the von Neumann entropy. In particular, the entropy production
is given by
\begin{eqnarray}
\frac{\rd S}{\rd t} = - 2 \lambda_k {\rm cov}({\hat\Phi}^k,\ln{\hat\rho}),
\label{eq:19}
\end{eqnarray}
where
\begin{eqnarray}
{\rm cov}({\hat X},{\hat Y})={\rm tr}({\hat\rho}{\hat X}
{\hat Y})-{\rm tr}({\hat\rho}{\hat X}){\rm tr}({\hat\rho}{\hat Y}).
\end{eqnarray}
The derivative of the entropy vanishes identically for pure
states, which is why the pure-state limit (\ref{eq:18}) is well
defined. In general, however, we see that $S$ is not necessarily constant. This is evident in the example
shown in Figure~\ref{fig:1}.  On the other hand, just as in
(\ref{eq:8}), the evolution equation (\ref{eq:17}) is autonomous and independent of the specific composition of the
mixture. Whether the fact that the entropy is variable raises
an issue remains an open question.

%\vskip 10pt The authors thank ... for comments and stimulating
%discussions. \vskip 10pt

\end{document}